\documentclass[sigconf,nonacm,screen]{acmart}

\usepackage{algorithm}
\usepackage{algpseudocode}
\usepackage{dsfont}

\usepackage{pgfplots}
\pgfplotsset{compat=1.17} 

\usepackage{microtype}
\usepackage{xcolor}
\usepackage{tabularx}
\usepackage{booktabs}
\usepackage{xspace}
\usepackage{balance}
\usepackage{multicol}
\usepackage{multirow}

\usepackage{nicefrac}
\usepackage{microtype}

\usepackage{balance}

\usepackage{paralist}

\AtBeginDocument{%
  \providecommand\BibTeX{{%
    \normalfont B\kern-0.5em{\scshape i\kern-0.25em b}\kern-0.8em\TeX}}}
\setcopyright{none}

\newcommand{\mA}{\mathbf{A}} 
\newcommand{\mC}{\mathbf{C}}

\newcommand{\mS}{\mathbf{S}}
\newcommand{\transpose} {^{\mbox{\scriptsize \sffamily T}}}

\newcommand{\xdrop}{$X$-Drop\xspace%
}
\newcommand{\PASTIS}{PASTIS\xspace%
}
\newcommand{\ELBA}{ELBA\xspace%
}
\newcommand{\ignore}[1] {%
}

\newcommand{\seqan}{SeqAn\xspace%
}

\newcommand{\ECOLI}{\emph{E. coli}\xspace}
\newcommand{\ECOLIX}{\emph{E. coli 100x}\xspace}
\newcommand{\CELEGANS}{\emph{C. elegans}\xspace}

\begin{document}
\title{Space Efficient Sequence Alignment for SRAM-Based Computing: X-Drop on the Graphcore IPU}

\author{Luk Burchard}
\email{luk@simula.no}
\authornotemark[1]
\affiliation{%
  \institution{Simula Research Laboratory}
  \streetaddress{Kristian Augusts gate 23}
  \city{Oslo}
  \country{Norway}
  \postcode{0164}
}

\author{Max Xiaohang Zhao}
\email{max.zhao@charite.de}
\authornote{Both authors contributed equally to this research.}
\affiliation{%
  \institution{Charité Universitätsmedizin}
  \streetaddress{Charitépl. 1}
  \city{Berlin}
  \country{Germany}
  \postcode{10117}
}
\author{Johannes Langguth}
\email{langguth@simula.no}
\affiliation{%
  \institution{Simula Research Laboratory}
  \streetaddress{Kristian Augusts gate 23}
  \city{Oslo}
  \country{Norway}
  \postcode{0164}
}

\author{Aydın Buluç}
\email{abuluc@lbl.gov}
\affiliation{%
  \institution{Lawrence Berkeley National Laboratory}
  \streetaddress{1 Cyclotron Rd}
  \city{Berkeley}
  \state{CA}
  \country{USA}
  \postcode{94720}
}

\author{Giulia Guidi}
\email{gg434@cornell.edu}
\affiliation{%
  \institution{Cornell University}
  \streetaddress{107 Hoy Rd City}
  \city{Ithaca}
  \state{NY}
  \country{USA}
  \postcode{14853}
}

\renewcommand{\shortauthors}{Burchard, et al.}

\begin{abstract}
Dedicated accelerator hardware has become essential for processing AI-based workloads, leading to the rise of novel accelerator architectures.
Furthermore, fundamental differences in memory architecture and parallelism have made these accelerators targets for scientific computing.

The sequence alignment problem is fundamental in bioinformatics; we have implemented the \xdrop algorithm, a heuristic method for pairwise alignment that reduces search space, on the Graphcore Intelligence Processor Unit (IPU) accelerator.
The \xdrop algorithm has an irregular computational pattern, which makes it difficult to accelerate due to load balancing.

Here, we introduce a graph-based partitioning and queue-based batch system to improve load balancing.
Our implementation achieves $10\times$ speedup over a state-of-the-art GPU implementation and up to $4.65\times$ compared to CPU.
In addition, we introduce a memory-restricted \xdrop algorithm that reduces memory footprint by $55\times$ and efficiently uses the IPU's limited low-latency SRAM.
This optimization further improves the strong scaling performance by $3.6\times$.

\end{abstract}
\maketitle

\vspace{-0.8em}
\section{Introduction}
\newcommand{\XDrop}{$X$-Drop\xspace}
Today's architectures are complex but often suboptimal for modern irregular computation, being overprovisioned for arithmetic computation and challenging the programmer to cope with the high cost of moving data.
A clear insight into this problem is provided by the Top500 list, in which the world's 10 fastest machines achieve peak performance of up to $83\%$ in the computationally intensive LINPACK benchmark but no more than $3\%$ peak performance in the High-Performance Conjugate Gradient (HPCG) benchmark, which involves irregular computation~\cite{meuer2001top500}.

In the last decade, the Graphics Processing Unit (GPU) has emerged as a leading architecture for high-performance computing (HPC) challenges involving dense linear algebra and scientific computing~\cite{vestias2014trends}.
However, GPUs are single instruction multiple data (SIMD) architectures, and this can be a limitation for computational challenges that suffer from high load imbalance, as is often the case with data analytics and general computation.
They require regular data access and work pattern to reach their theoretical peak performance~\cite{lee2010debunking}.
A general-purpose processor such as a CPU is better suited for non-uniform data access but does not provide the high instruction throughput achieved by GPUs, since CPUs are optimized for latency rather than throughput.
Therefore, new architectures are needed for HPC that can provide more flexible acceleration like CPUs while providing high throughput like GPUs.

%
Recently, the Graphcore Intelligence Processing Unit (IPU), a massively parallel multiple instruction multiple data (MIMD) SRAM-based processor designed as an AI accelerator, has emerged as a potential solution to irregular computation by combining fine-grained memory access with wide parallelism~\cite{jia2019dissecting}.
While processors connected to external RAM are constrained by the von Neumann bottleneck, SRAM-based computing eschews complex memory hierarchies by providing sufficient SRAM storage on the processing chip to fit a problem instance~\cite{mittal2021survey}.
IPUs were developed for AI applications but showed potential for other applications, such as for the breadth-first search algorithm, stencil computations, and cardiac simulation~\cite{burchard2021ipug,burchard2023enabling,louw2021using}.
The question we seek to answer in this work is if new SRAM-based architectures can improve performance on a wider range of emerging HPC challenges, such as bioinformatics.

\begin{figure}[htb]
   \centering
   \includegraphics[width=.75\columnwidth]{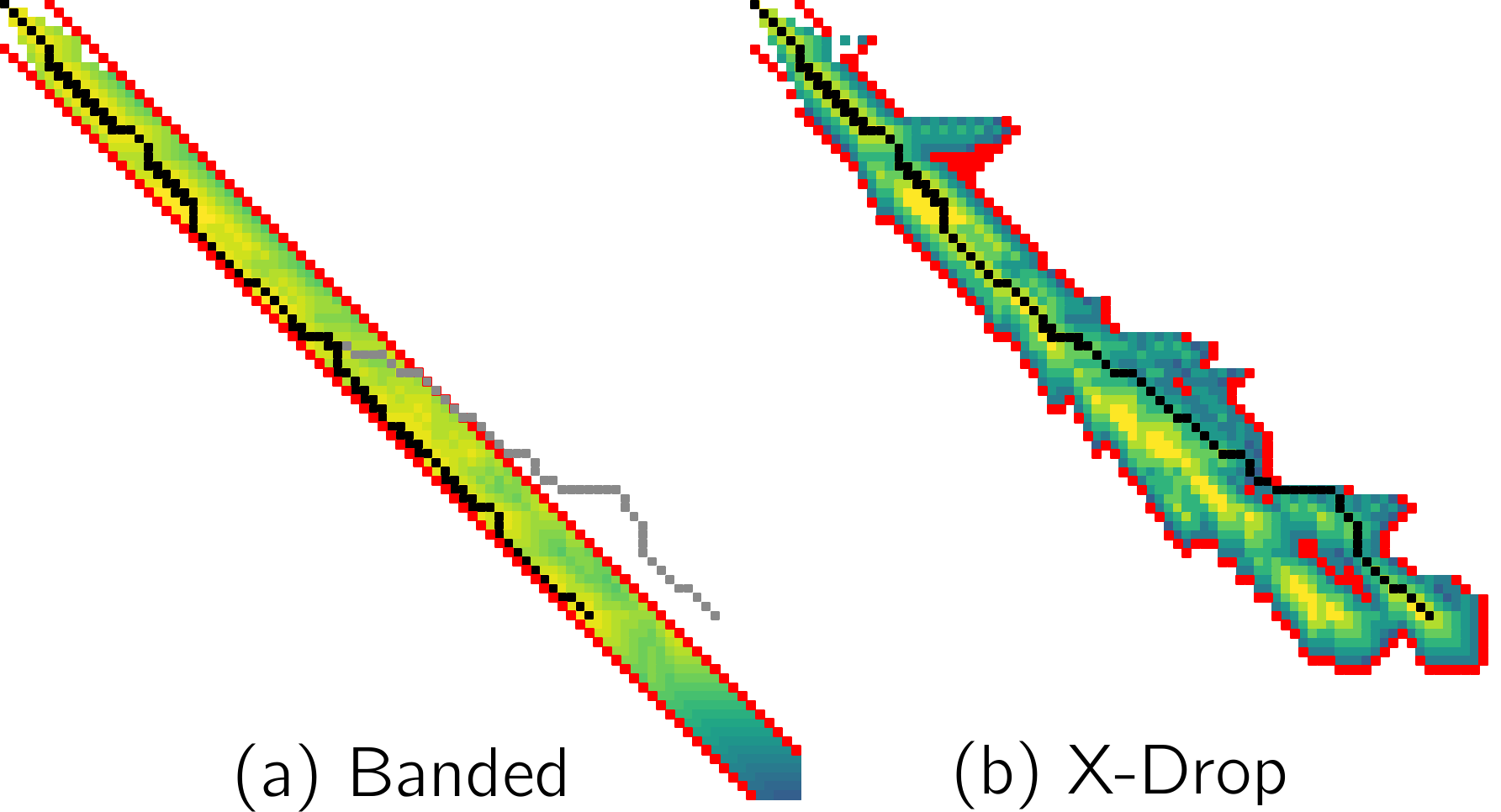}
   \caption{
   \normalfont
       On the left, the alignment is forced to stay within the banded area regardless of the score, missing the optimal alignment (gray). On the right, when the score (yellow-blue) $X$ goes below the current best score, the search is terminated (red boundary), and the optimal alignment (black) is returned.
    } 
   \label{fig:fig1}
   \vspace{-0.3cm}
\end{figure}

In bioinformatics, pairwise sequence alignment is a fundamental technique used in many scenarios, such as genome assembly, phylogenetic analysis, protein structure prediction based on homology, and searching for similar sequences in databases~\cite{apostolico1998sequence}.
Long-read sequencing technologies are increasingly available, producing sequences with an average length of about $15,000$-$30,000$ base pairs (bp).
Longer sequences allow for more precise genome assembly~\cite{amarasinghe2020opportunities} but they come with increased algorithmic complexity and computational cost.
Therefore, there is a need for efficient algorithms that can handle long-read sequencing data.

The Needleman-Wunsch (NW) algorithm is used for finding the best global alignment, while the Smith-Waterman (SW) algorithm is used for finding the best local alignment.
There is also a version of pairwise alignment called semi-global, where one side of the sequences is forced to align, but the other is not.
However, finding the optimal solution for these algorithms requires quadratic time as a function of sequence length, which is inefficient for long sequences.
The critical role that sequence alignment plays in understanding protein and DNA sequences has made it a focal point in attempts to optimize both algorithms and hardware~\cite{alser2021technology}.
In practice, assumptions can be made based on the input data and the type of computation desired to create heuristic algorithms with subquadratic runtimes.

A popular heuristic that we target in this work is called \xdrop.
The \xdrop algorithm is a heuristic for restricting the search space of a semi-global alignment algorithm.
It reduces the quadratic cost by dynamically searching only for a high-quality alignment and stopping the computation early when a good alignment is impossible. This allows for a more dynamic fit to the data than a static search space (Figure~\ref{fig:fig1}).
It is a promising algorithm for long-read sequencing data because a good alignment can be found in nearly linear time.
Genomic pipelines already use \xdrop and its variants $Y$-Drop and $Z$-Drop due to their good alignment quality and fast runtime~\cite{guidi2020Parallel, selvitopi2020distributed, altschul1990basic, li2018minimap2}.

In the literature, we only found one implementation of the \xdrop algorithm on GPUs~\cite{zeni2020logan}, and one on Field Programmable Gate Arrays (FPGAs)~\cite{zeni2021importance}, which can only run on DNA sequences (no protein sequences). However, \xdrop is widely implemented on CPUs~\cite{li2018minimap2, reinert2017seqan, altschul1990basic}.
In this paper, we present an implementation of the \xdrop algorithm on a novel AI accelerator hardware, the Graphcore IPU, that is suitable for both DNA and protein alignment.
The proposed IPU-based approach provides a competitive solution for accelerating \xdrop on a wider range of problem instances compared to traditional CPUs and GPUs. 
Here we demonstrate the practicality of our implementation by integrating it into two distributed-memory pipelines with high alignment volumes: ELBA, a \emph{de novo} long-read genome assembler~\cite{guidi2020Parallel,guidi2022distributed}, and PASTIS, a protein similarity search engine~\cite{selvitopi2020distributed}.
Our implementation is tested on a variety of real-world data, reporting speedup over  state-of-the-art CPU and GPU implementations.
Our work thus demonstrates the potential of AI architectures to accelerate the pairwise alignment of long sequences and their suitability for irregular scientific computations. 

IPUs and other SRAM-based devices, such as Cerebras hardware~\cite{lauterbach2021path}, rely on the MIMD paradigm and a large SRAM, making them more effective for handling irregular computation.
In this work, we not only highlight the advantages of SRAM-based computing for scientific computing and the speedup achieved for the \xdrop algorithm but also show where these novel devices need progress to be widely deployed.
For example, improving the IPU interconnect would lead to significantly cheaper host-to-device transfer times.

Our contributions are as follows:
\begin{compactitem}
    \item First, we demonstrate the first application of a cluster of Graphcore IPUs for high-performance processing of irregular genomic data.
    \item Then, we present a memory-restricted version of  \xdrop which reduces the required memory by a factor of up to $55\times$. This enables the algorithm to run in the IPU's SRAM memory.
    \item To solve the host-device communication bottleneck, to the best of our knowledge, we are the first to treat sequence comparisons as a graph and perform graph partitioning to reduce data transfer.
    \item Finally, we integrate our algorithm into ELBA and PASTIS, two state-of-the-art bioprocessing pipelines.
\end{compactitem}

\section{Background}\label{sec:background}
In this section, we first describe the architectural features of the Graphcore IPU and give a definition of the \xdrop pairwise alignment problem.
Then, we briefly describe two biological pipelines that use x-drop alignment and into which we have integrated our approach.

\subsection{Graphcore IPU}




The Graphcore IPU is a massively parallel multiple instruction multiple data (MIMD) processor consisting of a large number of independent units called \textit{tiles}.
Each of these tiles has a core and a small amount of SRAM memory.
Rather than serving as a cache, the SRAM memories of the individual tiles together form the device memory, eliminating the need for a traditional cache hierarchy.


\subsubsection{Hardware}
Three IPU generations have been released so far, called GC2, GC200, and BOW.
The BOW differs from the GC200 only in its clock frequency.
In this work, we use both the GC200 and the BOW.
Both IPU models, the GC200 and the BOW, consist of $1472$ tiles, each containing a core and $624$~KB SRAM, which run at $1.33$~GHz on the GC200 and at $1.85$~GHz on the BOW. 
Each IPU core runs six concurrent threads in \textit{temporal multithreading}, meaning that they are scheduled consecutively in a fixed order.
The majority of IPU instructions, including loading and storing from local tile memory, take exactly six cycles.
Thus, the $8832$ threads can be considered independent cores running at one-sixth the original clock frequency with no instruction or memory latency.
The total SRAM per IPU is $918$~MB, which can be read with an aggregate memory bandwidth of $46.9$~TB /s (GC200) or $65.2$~ TB/s (BOW).
However, data not local to a core must be communicated between tiles via the IPU exchange network, which has an aggregate bandwidth of $7.83$~ TB /s (GC200) or $10.9$~TB/s (BOW).
The IPU alternates between computation and communication in a bulk-synchronous parallel (BSP)~\cite{valiant1990bridging} manner with no overlap between phases.


Since the IPU is an accelerator that does not run its own operating system, it is dependent on a host machine.
Unlike GPUs, whose CPU host is usually within the same machine, a group of IPUs is connected to the host node via $100$~Gb/s Ethernet.
This means that the number of IPUs per host can vary greatly.

Our GC200 test system contains 64 IPUs, but only one dual-socket Xeon-based server.
Consequently, host-to-device transfers can become a bottleneck.
Four IPUs are used together in an IPU-M2000 blade, which in turn can be combined into larger systems called IPU-PODs.
The M2000 also contains up to $448$~GB of DRAM memory, which it can access at a rate of about $20$~GB/s.
This memory, while too slow for most computations, can be used to buffer data from host-to-device transfers. 
The IPUs themselves are connected in a ladder topology with a bisection bandwidth of $128$~GB/s.
In terms of power consumption, two IPUs are comparable to a powerful GPU like the NVIDIA A100 or a pair of moderately powerful CPUs~\cite{knowles2021graphcore}.


%

\subsubsection{Programmability}

Unlike other hardware accelerators, the IPU is a distributed memory system consisting of multiple tiles that use a direct memory write technique for communication.

Poplar is the C++ framework used to program the IPU at the lowest level.
It is inspired by TensorFlow and the dataflow programming model, as high-level program flow is defined as a dataflow graph, where we can define a state as a \textit{Tensor} and a transfer function as a \textit{Vertex}. 
Code that executes in a vertex is called a codelet.
One can think of a codelet as analogous to a CUDA kernel.
To synchronize computation and data accesses, the IPU hardware supports the BSP programming model, which divides algorithm execution into level-synchronous supersteps with three phases: Compute, Exchange, and Synchronize.

In Poplar, each \textit{Tensor} and \textit{Vertex} must be mapped to a tile.
The programmer must define input and output \textit{Tensor} for the \textit{Vertex}.
The compiler uses the dataflow graph and vertex mapping to create a synchronized data exchange following the BSP pattern.
A higher-level control flow can be introduced to select the next BSP superstep to execute.
Unlike MPI, the data exchange does not need to be explicitly programmed.

\subsection{X-Drop Pairwise Alignment}\label{sec:xdrop-pairwise-alignment}



The comparison of biological sequences is important for a deeper understanding of the role and function of genetic areas and protein structures, but also for the construction of the genomic sequence itself.
The genome consists of strings of nucleotides (adenine, thymine, guanine, cytosine), which code for protein sequences and contain additional regulatory information.
Genomes cannot be sequenced in their entire length; current sequencing technologies can only read and output sequences that are significantly shorter than the entire genome.
Therefore, we need sequence alignment to reconstruct whole genomes.
For short-read technologies such as Illumina, the average sequence length is 100-250 nucleotides (or base pairs, bp).
In newer long-read technologies such as Pacific Biosciences and Oxford Nanopore, the average read length can be more than 20,000 bp and up to several megabases, enabling the generation of highly continuous bacterial genomes~\cite{sereika2022oxford}.
Long-read technologies are highly promising as they can further improve our understanding of genomic structure~\cite{nurk2022complete}.
Yet, they also present new computational challenges due to their longer length and higher error rates.

The optimal sequence alignment between two sequences can be found in quadratic time and linear space~\cite{hirschberg1975linear,myers1988optimal} if we use the classical Smith-Waterman or Needleman-Wunsch algorithm for local and global alignment, respectively. 
The sequence alignment problem is defined as follows.
Given two sequences $\mathcal{H} = h_1, h_2, \ldots, h_m $, $\mathcal{V} = v_1, v_2, \ldots, v_n$, with $|\mathcal{H}| = m, |\mathcal{V}| = n$ we want to find the best scoring set of changes to transform sequence $\mathcal{H}$ into $\mathcal{V}$. 
If we assume that the sequences are homologous, i.e. that they are evolutionarily related, the number of resulting changes is small.
The alignment is done by dynamic programming, where we define a dense scoring matrix $S(i, j)$, with $i \leq n, j \leq m$.
The matrix $S$ is filled from the upper left corner and extended to the lower right corner.
In each nonzero, we store the best score for the alignment of each two-symbol pair $(v_i, h_j)$.
This score is computed based on the match of $v_i$ and $h_j$ and the history of alignment for the previous three scores, as defined in the following rule:
$$ S(i, j) = \begin{cases}
    S(i-1, j-1)+Sim(v_i, h_j) & \text { if } i>0, j>0 , \\
    S(i, j-1) +    \text{gap} & \text { if } j>0 \\
    S(i-1, j) +    \text{gap} & \text { if } i>0 \\ \end{cases} $$
In the above definition, $Sim(v_i, h_j)$ is an arbitrary scoring function used to quantify the degree of similarity between a pair.
In the case of DNA, $Sim(v_i, h_j)$ is a positive value if $v_i$ and $h_j$ match (i.e., no change is required), or a negative value if they do not, and gap is also a negative value, meaning that either $v_i$ or $h_j$ has a symbol inserted or deleted at that position:
The goal is to find a path of changes in $S$ that maximizes the score and is optimal for aligning $\mathcal{H}$ and $\mathcal{V}$.
The scoring function assigns higher scores to likely biologically related sequences and lower scores to less likely related sequences. 

In real-world scenarios, we can often make a reasonable assumption about where to find the optimal alignment on the two sequences, and this can lead to heuristics that can significantly reduce time and space complexity. 

\begin{figure}
    \centering
    \includegraphics[width=\linewidth]{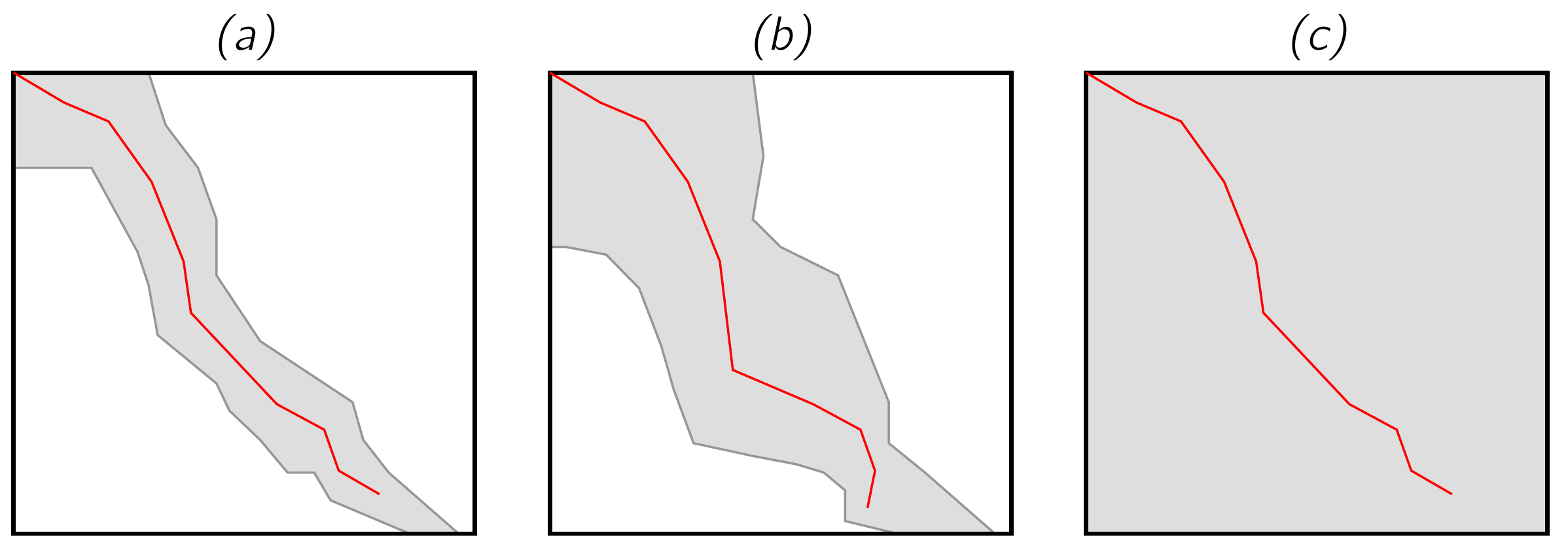}
    \vspace{-0.7cm}
    \caption{
    \normalfont 
    The red path is the optimal alignment, the gray area is calculated values, and the white area is non-calculated values.
    Due to the \xdrop condition, the white nonzeros contain a score of $-\infty$.
    Panel (a) shows an iteration with $X = 10$, (b) with $X = 20$, and (c) with $X=\infty$.
    }
    \label{fig:xdrop_effect}
\end{figure}

First, in one-to-many and many-to-many sequence alignment, the number of sequences to be compared can be reduced by first identifying common contiguous subsequences of fixed length $k$ (i.e., $k$-mers).
The $k$-mer information reduces the number of sequences to be compared in large-scale computation but also gives us an indication of where in the sequences the optimal alignment might be found.
This can lead to a semi-global alignment approach, where each pairwise alignment is divided into the left and right extension of the $k$-mer match.

The semi-global approach forces the alignment to start at one of the two extremities of the sequences (similar to the Needleman-Wunsch algorithm~\cite{needleman1970general}), but leaves the other extremity free.
This is a common approach first introduced by BLAST ~\cite{altschul1990basic}, and it can lead to shorter sequences, but it scales with $\mathcal{O}(mn)$, where $m$ and $n$ are the lengths of the sequences involved since we still need to compute the entire matrix $S$.

The second key insight is that for sequences that have some similarity, the optimal alignment is often found on the diagonal of the $S$ matrix, with the antidiagonal extremities storing low (i.e., bad) scores because the number of mismatches is high when moving away from the center of the diagonal, as shown in Figure~\ref{fig:fig1}.
A common approach is to restrict the search to a predefined band region of $S$ around the diagonal (left in Figure~\ref{fig:fig1}).
This heuristic significantly reduces the time and space complexity but may restrict the search space too much and result in the failure to find the correct optimal alignment.
This problem is particularly severe for sequences generated with long-read technologies, as these technologies are more prone to insertion and deletion of nucleotides (which can lead to a long gap sequence that moves the optimal alignment away from the diagonal) than to mismatches (which keep the optimal alignment mostly on the diagonal), which was the case with short-read technologies.

This provides the motivation for the \xdrop condition.
The \xdrop strategy can be viewed as a dynamic band approach, where the search space is dynamically bounded by the score values rather than by a predefined band width. The \xdrop algorithm defines a threshold $X$ that removes nonzeros (and the resulting possible path) from the $S$ matrix that are worse than a path with the current best score.
The assumption is that a path that is significantly worse (where significant is defined by $X$) than the current best score is unlikely to lead to the optimal alignment.
Let us denote the current best score as $T$.
The \xdrop condition states that if $S(i, j) < T - X$ in the current iteration of the dynamic program, then $S(i, j) = -\infty$.

The first iteration of the fill process for the \xdrop algorithm initiates in the upper left corner and initializes the matrix with $S(0, 0) = 0$.
To avoid starting the alignment with a gap or gap sequence at a location in the matrix other than the upper left corner (i.e., where both sequences start), we define the off- matrix access as $S(i, j) = -\infty, i < 0 \vee j< 0$.
This is because we want to perform a semi-global alignment and force one side of the extremities (from the $k$-mer match heuristic) to align.
If $X$ is large, the computation approaches filling the entire dynamic programming matrix $S$, whereas when $X$ is small, the non-zeros in $S$ are pruned as we move away from the optimal result, and these non-zeros become $-\infty$.
In Figure~\ref{fig:xdrop_effect} we show the impact of different $X$ values on the search space of the scoring matrix.


The \xdrop algorithm has been widely implemented and variants used for commonly used long-read alignment software such as minimap2~\cite{suzuki2018introducing}.
In this work, we implemented the original \xdrop algorithm as formulated by Zhang~\cite{zhang1998alignments, zhang2000greedy}. 
Their implementation traverses the dynamic programming matrix $S$ in an antidiagonal fashion.
They fill the non-zeros along the antidiagonal line from the upper right corner to the lower left corner of the dynamic programming matrix $S$.
This sweeping anti-diagonal approach is initiated at $S(0,0)$ and progresses until the lower right corner is reached.
To fill a cell in an antidiagonal of the matrix, only the scores of the adjacent cells (top, top-left, and left) are needed.
These are stored in the antidiagonals that were filled in the previous two phases.
Earlier literature~\cite{zhang1998alignments, zhang2000greedy} used this insight to store only three antidiagonals: two for the previous phases and one for the current phase.
This approach is popular for SIMD parallelism because the dependencies for input and output non-zeros are well aligned.
But this is not strictly necessary. Gotoh~\cite{gotoh1982improved} stated that storing two antidiagonals is sufficient. However, it is not often used in practice because SIMD parallelism is difficult to achieve.
Despite that, in this work, we choose to store only two antidiagonals as the IPU is a MIMD architecture, and we aim to reduce the memory footprint. 




\subsection{ELBA}\label{sec:elba}

ELBA is a long-read assembler implemented for distributed-memory parallelism that uses sparse matrices as the main data structures, mapping the \emph{de novo} assembly process onto sparse matrix computation~\cite{guidi2022distributed}.
ELBA is composed of five main stages which comply with the Overlap-Layout-Consensus paradigm for assembling long-read sequencing data.
In the first step, $k$-mer counting, the input sequences are parsed to extract subsequences of fixed length $k$ and count their frequency. This produces a 1D distributed hash table of the $k$-mers and their frequencies and sequences of origin.
This hash table is then transformed into a 2D $|k-mers|$-by-$|sequences|$ sparse matrix that we call $\mA\transpose$. 
In the second step, called overlap detection, ELBA multiplies $\mA$ by its transpose $\mA\transpose$ to detect overlapping $k$-mer matches between input sequences. 
In this way, a $|sequences|$-by-$|sequences|$ matrix $\mC$ is obtained in which the non-zeros represent such matches and their position on the sequences.
Then, for each non-zero of $\mC$, the \xdrop pairwise alignment algorithm is run, starting from the $k$-mer march position, to obtain similarity values and remove false matches from the matrix.
In the fourth and fifth stages, ELBA simplifies the matrix, that is the assembly graph, to extract contiguous areas of the genomes (i.e., \emph{contigs}), which are the result of the assembly process.


\subsection{PASTIS}\label{sec:pastis}

PASTIS~\cite{selvitopi2020distributed} similar to ELBA computes protein homology searches as a distributed sparse matrix multiplication.  
PASTIS computes the $k$-mer count and $\mA\mA\transpose$, but then must perform additional matrix multiplication with the matrix $S$ to produce the output $|sequences|$-by-$|sequences|$ matrix.
The $S$ matrix is called the substitution matrix and is used to find quasi-exact $k$-mer matches because it has been shown that strictly enforcing exact matches in protein homology searches can lead to a significant loss of accuracy.
Thus, the overlap detection phase has the form of $\mA\mS\mA\transpose$.
Once the output matrix is formed, PASTIS computes an alignment step on each non-zero, similar to ELBA.
PASTIS has two alignment modes: seed-and-extend with \xdrop and Smith-Waterman alignment.
Using \xdrop, PASTIS initiates the alignment from the $k$-mer match.

Both PASTIS and ELBA defer implementation of \xdrop to the Library for Sequence Analysis (\seqan) C++ library for CPU ~\cite{reinert2017seqan}.
ELBA also provides support for the GPU-based \xdrop alignment called LOGAN ~\cite{zeni2020logan}.
LOGAN does not support protein alignment.

\section{Algorithm}\label{sec:algorithm}
In this section, we describe the algorithm we implemented on the Graphcore IPU and the algorithmic changes we made to make the computation more suitable for the IPU.

\begin{figure}
    \centering
    \includegraphics[width=\linewidth]{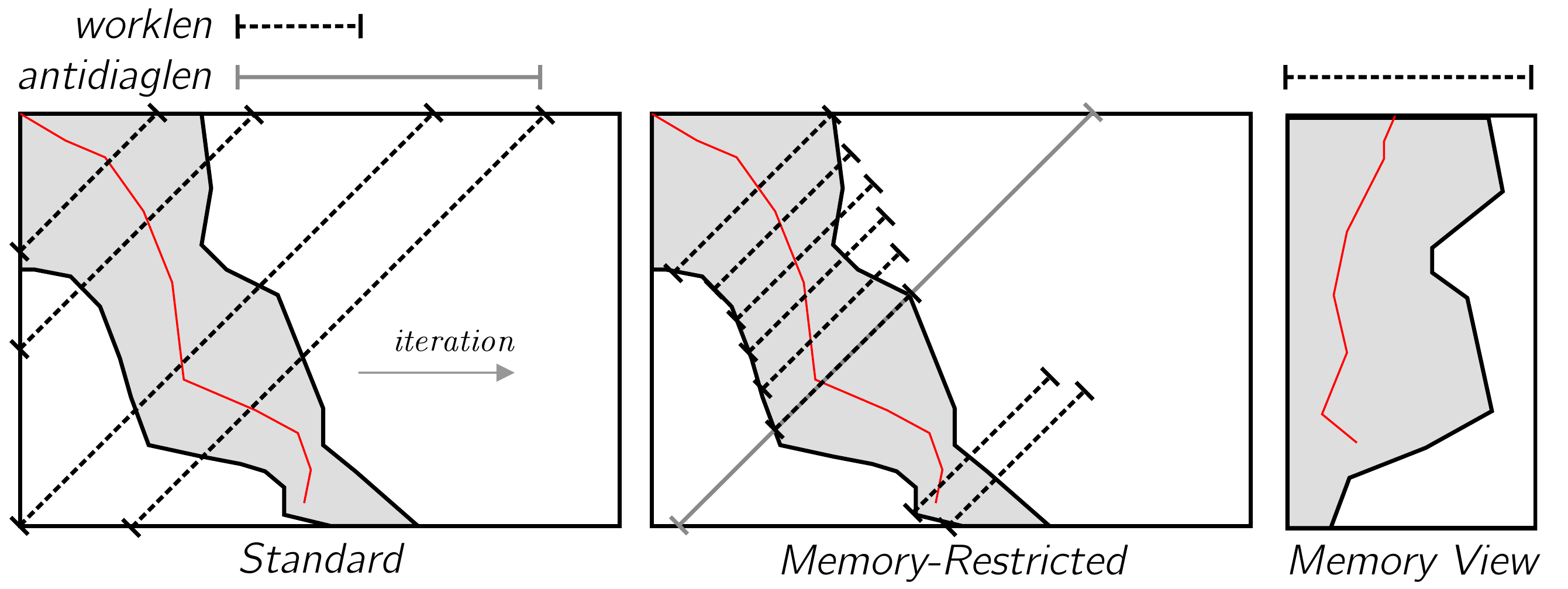}
    \vspace{-0.7cm}
    \caption{{
    \normalfont 
    The antidiagonal length is $\delta = min(|\mathcal{H}|, |\mathcal{V}|)$.
    The memory-restricted version allocates work memory of $\max_{k} |U_k-L_k| \leq \delta_b \leq \delta$. The left panel illustrates the standard algorithm ($3\delta$ memory).
    The middle one illustrates our algorithm ($2\delta_b$ memory), while the right one is the pattern of our memory usage over time.
    }}
    \label{fig:bandalgo}
\end{figure}

One of the major challenges in implementing sequence alignment on specialized hardware is the memory requirement since storing the entire dynamic matrix can exceed the available memory.
In Section~\ref{sec:xdrop-pairwise-alignment}, we described how it is possible to reduce the memory footprint of the scoring matrix $S$ by storing only three antidiagonal phases (the previous two phases and the current phase $k$) to traverse $S$, instead of storing the entire matrix.
It can be observed that an antidiagonal can never become larger than $\delta = min(|\mathcal{H}|, |\mathcal{V}|)$, where $\mathcal{H}$ and $\mathcal{V}$ are the sequences involved in the alignment.
To limit the workload, it is common to use a lower $L_k$ and an upper bound $U_k$ for the antidiagonal, where $|U_k - L_k|$ is the length of the antidiagonal in iteration $k$.
These boundaries are derived from the number of scores in $S$ that are not yet $-\infty$ (i.e., scores that have not triggered the \xdrop termination condition). 

To store three antidiagonal phases, we need $3\delta$ of memory for each alignment run.
This memory requirement is too high for the IPU. Therefore, we address this problem with a two-step approach.
First, we reformulate the algorithm using the technique found in~\cite{gotoh1982improved} to store only two antidiagonal phases.
This is possible by using a temporary variable since the values in the antidiagonal $k$ and $k-2$ are one iteration offset accessed and written.
In addition, we propose to use a band in the iteration, which is different from the classical banded algorithm shown in Figure~\ref{fig:fig1} on the left, because the band is not static in space (i.e., it does not remain fixed around the diagonal), but is constantly realigned to the active iteration position that stores the best score.
It is possible to observe that even though the antidiagonal is fully allocated ($\delta$), only a small part of it is accessed during each phase $k$, since $|U_k - L_k| \leq \delta$. Therefore, in our implementation, we assume a bound length $\delta_b$, which is the total working length $w = \max_{k} |U_k-L_k|$ of the antidiagonal to keep $w \leq \delta_b \leq \delta$.
Thus, we use the restricted $\delta_b$ to constrain the algorithm in memory by placing antidiagonals in the active working area of the algorithm, resulting in a memory allocation of $2\delta_b$.
Figure~\ref{fig:bandalgo} on the left illustrates the antidiagonal length (black dashed line) for the original algorithm, while the middle one illustrates the antidiagonal length for our proposed memory-restricted version.
The gray area is part of the scoring matrix $S$ filled by the \xdrop algorithm.
The right panel in Figure~\ref{fig:bandalgo} illustrates a reinterpretation of the iteration space of the working memory region.
The choice of an appropriate $\delta_b$ value is related to the error rate of the sequence and the \xdrop factor.
Both high error rates and large $X$ increase the working length $w$, as shown in Section~\ref{sec:delta}.

Algorithm~\ref{alg:xdropdoubleband} describes our memory-restricted algorithm using only two antidiagonals $A_1, A_2$ of length $\delta_b$.
$T$ is the best score found by the algorithm, while $L, U$ are the lower and upper iteration boundaries, respectively.
It is worth remembering that the algorithm for aligning \xdrop is semi-global; one side of the extremities of the two sequences is forced to align while the other side is left free.
This is the case because each alignment results from splitting two sequences into four sequences (i.e., two for the left extension and two for the right extension) using the $k$-mer seed match information.
To perform the forward alignment (right extension), we can access the sequences in a natural access pattern from left to right.
For the backward alignment (left extension), we use an index transformation $op(\cdot)$ that produces either forward or backward accesses to $\mathcal{H}$ and $\mathcal{V}$.
This way, we do not have to completely reverse the sequences to perform the alignment with the left extension.
The current diagonal iteration is given by $k$.
The algorithm terminates when $L$ and $U$ converge, i.e., when no values greater than $-\infty$ remain in the working set of the algorithm.


\begin{algorithm}
\caption{The memory-restricted \xdrop algorithm.}\label{alg:xdropdoubleband}
\begin{algorithmic}[1]
\State $L, U, T', T, k \gets 0$
\State $L1_{inc}, L2_{inc} \gets 0$
\State $A_1, A_2 \gets \{-\infty, \ldots, - \infty\}$
\State $A_1[0] \gets 0$
\While{$L \leq U + 1$ increase $k$ by $1$ and}
\State $W_2 \gets A_2 + (-L + L2_{inc})$ \Comment{C-style array offsetting}
\State $W_1 \gets A_1 + (-L + L2_{inc} + L1_{inc})$
\State $W_1' \gets A_1 + (-L)$

\State $w_{last} \gets W_1[L - 1]$ \Comment{Instead of a third anti-diagonal}
\While{$i \in (L, \dots, U + 1)$}
\State $j \gets k - i - 1$
\State $w_{new} \gets W_1[i]$
\State $score \gets max \left\{ 
 \begin{array}{ r }
  W_2[i] - gap \\
  W_2[i - 1] - gap \\
  w_{last} + sim(\mathcal{H}[op(i)], \mathcal{V}[op(j)])) \\
\end{array}
\right\}$
\State $w_{last} \gets w_{new}$
      \If{$score < T - X$} 
        \State{$score \gets \xspace -- \infty$}
      \EndIf
\State $W_1'[i] \gets score$
\State $T' \gets max\{T', score\}$
\EndWhile
\State $L_{prev} \gets L$
\Comment{zero t1 shifted values}
\State $L \gets max(k + 1 - N, argmin(W_1' \neq - \infty))$
\State $U \gets min(|\mathcal{H}| - 1, argmax(W_1' \neq - \infty) + 1)$
\State $L1_{inc} \gets L - L_{prev}$
\State $T \gets T'$
\State $swap(A_1, A_2)$
\State $swap(L1_{inc}, L2_{inc})$
\EndWhile
\end{algorithmic}
\end{algorithm}

\section{Implementation}
In this section, we describe the implementation of the memory-restricted \xdrop algorithm on the Graphcore IPU accelerator.

Our implementation focuses on large sequences (both protein and DNA) whose length is in the range of $1$K to $25$K.
This raises two challenges that we address in this work.
First, the memory requirement for each alignment is large, given that a single pairwise alignment is executed on a single tile of the IPU, where that tile has $624$ KB of addressable memory and six threads, each of which requires space for the algorithm.
So we need to be able to allocate $6\times$ the amount of working memory during the alignment on one tile.
Second, due to the IPU's BSP (bulk synchronous parallel) architecture, we need to create a load-balanced problem (i.e., with equal runtime for each tile) to use all tiles equally.
If a single tile takes more time, all other tiles must wait, resulting in poor utilization of hardware resources.

\subsection{Kernel Architecture}

The \xdrop kernel was written as a Poplar vertex (codelet) in C++, where Poplar is the equivalent of CUDA for IPUs.
Our implementation focuses on using the six hardware threads of each IPU tile. 
In the absence of synchronization instructions such as atomics or mutexes, we implemented a data-parallel implementation with throughput in mind.
It is possible to perform tile-local coarse-grained synchronization by combining the hardware threads into a single supervised thread, which can use the entire set of six threads on a single alignment.
However, this would result in context switches for each synchronized part of the algorithm, which would degrade performance.
Therefore, we choose to have each thread perform a single performance alignment, and for this reason we have a sixfold memory footprint on each tile.

The IPUs inability for random external memory access and the limitation of a single tile to $624$~KB of local memory forces us to choose a memory minimizing algorithm that enables many input sequences to be stored on a tile in order to maximize the number and size of sequences that can be processed on a single tile.
To optimize the use of limited local memory, our tile architecture employs several techniques that allow all six threads to be used during the execution of the \xdrop algorithm.
By reducing the amount of memory needed to implement the memory-restricted algorithm and efficiently using the available processing resources, we can achieve better performance and efficiency in sequence alignment on the IPU.
The optimizations we implemented and their relative improvement are summarized in Table~\ref{tab:improvement}, measured for real-world and synthetic data.


\begin{table}[]
\caption{
\normalfont
Optimizations implemented and described throughout Section~\ref{sec:impl}.}
\vspace{-0.4cm}
\label{tab:improvement}

\begin{tabularx}{\linewidth}{@{}llrrrr@{}}
\cmidrule(l){2-6}
                                                         & Optimization        & Time [ms]    & GCUPS     & To Prev.        & Total    \\ \cmidrule(l){1-6} 
\multirow{6}{*}{\rotatebox[origin=c]{90}{$15\%$ error}}  & Single tile         & $493907 $    & $5.00 $   &                 &          \\
                                                         & Scale to 1472 tiles & $414    $    & $6034 $   & $1194\times$    & $1193.8\times$ \\
                                                         & Use 6 threads       & $87     $    & $28705$   & $4.76\times   $ & $5679.4\times$ \\
                                                         & LR splitting        & $85     $    & $29163$   & $1.02\times   $ & $5768.0\times$ \\
                                                         & Work-stealing       & $85     $    & $29084$   & $1.00\times   $ & $5765.8\times$ \\
                                                         & Dual issue          & $65     $    & $37933$   & $1.30\times   $ & $7504.9\times$ \\ \cmidrule(l){1-6} 
\multirow{6}{*}{\rotatebox[origin=c]{90}{ELBA Ecoli}}    & Single tile         & $4180499$    & $14.52$   &          &          \\
                                                        & Scale to 1472 tiles  & $6939   $    & $7302 $   & $602\times$     & $ 602.4\times $ \\
                                                        & Use 6 threads        & $2707   $    & $14860$   & $2.56\times  $  & $1543.9\times$ \\
                                                        & LR splitting         & $2470   $    & $16828$   & $1.10\times  $  & $1692.3\times$ \\
                                                        & Work-stealing        & $1713   $    & $21935$   & $1.44\times  $  & $2440.4\times$ \\
                                                        & Dual issue           & $1268   $    & $28587$   & $1.35\times  $  & $3296.8\times$ \\ \cmidrule(l){1-6} 
\end{tabularx}
\end{table}

\subsubsection{Tile Data Structures} 

\begin{figure}
    \centering
    \includegraphics[width=\linewidth]{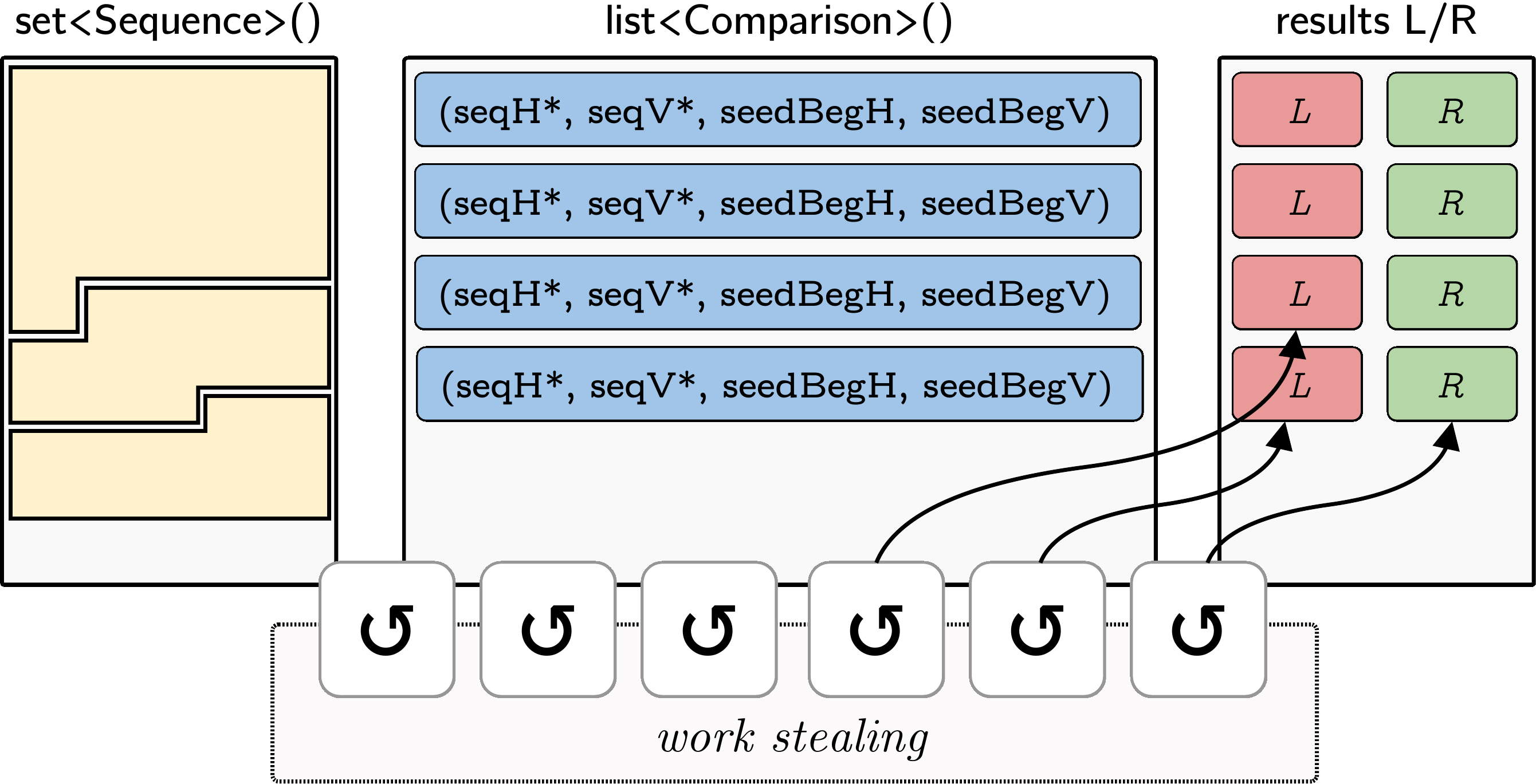}
    \caption{
    \normalfont Tile structure with six worker threads filling in the output for left and right seed extension, using work stealing. The input is sequences and a list of seed extension information.
    }
    \label{fig:tilestructure}
\end{figure}

The tile receives as input a set of sequences \texttt{seqs} and a list of seeds for these sequences to be computed.
The seed matches are tuples containing a pointer to two sequences in \texttt{seqs}, and the position of the seeds on them to avoid having to split the sequences on the host, as shown in Figure~\ref{fig:tilestructure}.
The output array stores a list of tuples for each left and right extension of a seed.
Our representation has many advantages over state-of-the-art seed extension representation, as no preprocessing has to be done on the host device. 
Our algorithm can operate in reverse on continuous memory by providing a suitable $op$ function for Algorithm~\ref{alg:xdropdoubleband}.
Furthermore, in real-world pipelines, a pair of sequences often must be aligned considering multiple seed matches, which would lead to the retransmission of the same sequences, negatively affecting performance.
Thanks to the detached structure for sequences and seed alignment information we introduced, we can transfer multiple seed matches and sequences at once.
This optimization saves $\mathcal{O}(\#seeds)$ in data transfer from the host to the device.
The use of the $op$ function is useful because the sequences can be truncated at different positions for the left and right extension, creating $\#matches\times4$ $\mathcal{H}_L, \mathcal{H}_R, \mathcal{V}_L, \mathcal{V}_R$ individual sequences.

\subsubsection{Left and Right (LR) Extension Splitting} 
Scaling from one thread to six threads per tile, we expect a speedup of $6\times$.
However, the observed speedup is only $4.7\times$.
This is because only $5$ comparisons with $10$ unique sequences have the memory to accommodate large sequences of length $10,000$~bp on a single tile, leaving one of six threads without work.
To use all threads, we introduce a finer distribution of work by having threads work individually on the left and right extensions of the seed matches rather than assigning both extensions to the same thread.
This doubles the number of work units to be distributed so that each thread is used for large sequences instead of leaving one thread idle.
Our synthetic data, whose sequences are generated to be of equal length, does not benefit from this optimization because if we have $5$ uniform workloads (i.e., sequence comparisons), they are split into $10$ uniform workload units.
This leaves four threads with two workload units, as is the case even without this optimization since we rely on BSP synchronization, which does not benefit from an unbalanced workload since the bottleneck is caused by the longest-running process.
Nevertheless, this optimization can lead to a significant improvement in real-world workload due to a larger variance in seed position and sequence length.

\subsubsection{Eventual Work Stealing}\label{sec:eventual-work-steal}

Despite the LR optimization to increase workload granularity, we can still observe a large variance in thread runtime due to sequence length variance in real data.
Using a simple round-robin workload allocation will leave one or more threads without work, resulting in load imbalance.
Since synchronization is not possible on the IPU tiles, except for coarse thread joining, we initially resorted to statically assigning work to individual threads.
However, since a single-seed extension has a relatively high runtime, we decided to implement \textit{Eventual Work Stealing}.
A work-stealing approach makes it possible for an idle thread to take a unit of work from the globally stored list of seed extensions and work on it locally.
Since no mutexes were available to ensure that only one thread at a time could access the seed structure, we resorted to globally swapping a value.
This does not avoid race conditions, but in this case, we would only compute a seed extension multiple times and not skip over it.
Since instruction latencies are deterministic, two threads stealing the same unit of work will perpetually continue to do so.
We introduced a small thread-unique busy wait loop to create variance, eventually avoiding this race condition and a possible perpetual joint execution. This loop reduces the race conditions from $16$K to $18$ for a total number of $1.13$M alignments performed.

\begin{table*}[tb]
	\caption{{\normalfont 
Data sets for comparisons with CPU and GPU implementations with distribution for the left and right extensions.
 }}
 \vspace{-0.4cm}
 \label{tab:datasets}
	\resizebox{\textwidth}{!}{%
 \begin{tabular}{llrrrrrrrrr}
\toprule
 Name & Cmp Count & Seqlen Avg & Seqlen P10 L & Seqlen Avg L & Seqlen P90 L & Seqlen P10 R & Seqlen Avg R & Seqlen P90 R & Complexity Avg\\
\midrule
simulated85 & $40\,000$ & $9\,992$ & $9\,992$ & $9\,992$ & $9\,992$ & $9\,991$ & $9\,991$ & $9\,991$ & $99\,830\,072$ \\
ecoli & $568\,208$ & $7\,319$ & $832$ & $7\,322$ & $13\,684$ & $823$ & $7\,317$ & $13\,675$ & $45\,870\,449$ \\
ecoli100 & $15\,611\,769$ & $3\,631$ & $431$ & $3\,705$ & $8\,319$ & $388$ & $3\,557$ & $8\,087$ & $12\,524\,999$ \\
elegans & $16\,794\,715$ & $7\,346$ & $1\,184$ & $7\,347$ & $13\,375$ & $1\,179$ & $7\,345$ & $13\,380$ & $52\,763\,834$ \\
\bottomrule
\end{tabular}
	}
\vspace{-11pt}

\end{table*}

\subsubsection{Dual Instruction Issuing} 

The tiles implement a Very Long Instruction Word (VLIW) ISA with dual instruction issuance on two lock-synchronous pipelines, one integer and one floating point.
The instructions require a single cycle to retire, except for certain floating-point operations such as \texttt{exp}, \texttt{log}, \texttt{sqrt}, which are not used in our code.
Thus, we are not concerned with lock synchronicity.
The integer pipeline is responsible for memory and branching operations, while the floating point pipeline is only responsible for floating point arithmetic.

We analyzed the generated assembly and found that registers in the integer pipeline in the inner loop of the \xdrop algorithm spilled when traversing the antidiagonal.
Therefore, we reformulated our similarity function $Sim$ (Section~\ref{sec:background}) to return floating-point values, forcing a floating-point representation of our scores to make use of additional floating-point registers.
In addition, we used a compiler hint to use the built-in floating-point max instruction.

\subsection{Batching}

Before the kernel is executed, the sequences must be distributed (in pairs) to the individual tiles.
This distribution can be modeled as a $k$-partitioning problem, where each comparison task is assigned to a particular tile and the total number of tiles is $k$. The size of each comparison is equal to the sum of the lengths of the two sequences involved. Given the BSP pattern of the IPU, it is important to minimize the longest-running tile runtime, which may cause other tiles to wait.

Estimating computational complexity is difficult because perfectly matched sequences have a smaller search space (they do not deviate too much from the diagonal) than sequences with higher mismatch rates.
However, completely mismatching sequences run faster because the computation terminates early due to the \xdrop condition triggered by rapidly increasing bad scores.
Therefore, we use the maximum running time, which is quadratic to the lengths of the sequences involved, for each comparison as an estimate of the computation time.


\subsection{Graph Based Sequence Partitioning}
\label{sec:graph-partitioning}

A single seed extension $e_c$ with comparison index $c$ for two sequences $\mathcal{H}, \mathcal{V} \in \Omega$, where $\Omega$ is the total set of input sequences, stores the following information $e_c := (\mathcal{H}, \mathcal{V}, \mathcal{H}_s, \mathcal{V}_s)_c, \in \mathcal{C}$, where $\mathcal{H}_s, \mathcal{V}_s$ refer to the initial position of the seed on the two sequences $\mathcal{H}, \mathcal{V}$, respectively.
In previous work, the relationship between pairs of sequences was not considered (e.g., when two identical sequences have multiple $k$-mer matches), but these $c$-tuples were considered as single sequence extensions to be computed.

In this work, we propose to interpret the set of seed extensions as a graph partitioning problem to reduce the number of data transfers between the host and the IPU.
The idea is to reduce the transmission of the same sequence of $\Omega$ shared by multiple $e_c$ sent to the IPU in the same batch.
Since memory is not shared between tiles and communication must be determined at compile time, we are limited to reusing sequences on a single tile.
Dynamic compilation of IPU exchanges at runtime takes too much time.
Therefore, we cannot create dynamic sequence exchanges and keep $\Omega$ entirely on the IPU.
Reuse of many sequences is enabled by our tile data structures, which store the local set of sequences $\omega_i \subset \Omega$ detached from the set of seed extensions in a tile with index $i$.
The set of seed extensions stores only a reference to the sequences in $\omega_i$.

Real-world bioinformatics pipelines, which rely on many-to-many sequence comparisons, store information about which sequences need to be aligned against each other, which can benefit our optimization. 
Both ELBA and PASTIS provide this information in a sparse matrix representation that includes planned sequence alignments.
Here, we propose a graph partitioning algorithm to increase data reuse.
Given a graph $G(V, E)$ with $V \subseteq \Omega$, $V$ the set of vertices (i.e., sequences), we want to distribute the set of edges $e_j \in E$ that is representative of the seed extensions (i.e., alignments) to be performed.
The graph has an edge between two vertices where a comparison uses the sequences represented in $V$.
We partition our graph into partitions $p_i$ containing a set of edges and the associated set of sequences unique to the tile.
The partitions are constrained to hold sequences whose total size is less than or equal to the available tile memory.

To avoid this immediate optimization step taking a lot of time, we partition the graph using a greedy strategy to stay within a tight time frame, which is usually less than one second for our tested data.
The greedy strategy is given as follows.
Take a vertex in the graph and walk linearly through the edge list.
Add the starting vertex to the partition and the adjacent vertex to the edge.
Continue to walk through the edges and add the adjacent vertex to the partition until adding a new vertex would exceed the memory limit of the partition; start a new partition.
For simplicity, we leave the batching of our partitions to our batching algorithm.

For sequences of the same length, this optimization achieves a sequence reuse effectiveness of $2\times$, since for each new comparison on a tile, only one new sequence has to be transmitted, since the other is already in $\omega_i$.
In a real world scenario, however, the length of the sequences can vary significantly.
Thus, for \ECOLI and \CELEGANS, we observed that we could pack up to $41$ smaller sequences into a single large sequence, which drastically improved the transfer performance.

\subsection{Multi-IPU Support}

The use of multiple IPUs is a critical factor in achieving optimal performance under a large computational load, as is often the case with many-to-many sequence alignment.
We have several options to scale our algorithm.
One of these options is the combined multi-device approach, which provides a virtual, seemingly homogeneously extended IPU with a larger number of tiles.
However, it should be noted that this approach can lead to global synchronization and requires larger batches, resulting in suboptimal parallelization efficiency.
Therefore, we opted for a different approach using multiple individual IPU devices.
With our load balancing driver, we can effectively manage load balancing and schedule alignments between connected standalone devices.
It is worth mentioning that the individual devices remain hidden from the user.
With such a method, we can achieve better performance while ensuring optimal resource utilization.

Our wrapping driver class manages the Poplar graph and enables execution on multiple IPUs.
The driver class handles the submission of batches and takes care of the internal distribution of work between IPUs and their respective tiles.
Batches are submitted to a work queue that is shared among all IPU instances.
The shared queue is connected to the input stream of each IPU, which allows prefetching by the IPU since all submitted batches are fully preprocessed.
Prefetching on the IPU allows data transmission to be interleaved with computation allowing us to hide transmission time to an extent.



\label{sec:impl}

\section{Experimental Setup}\label{sec:setup}
Our tests were performed on three large systems.
First, we used the Perlmutter supercomputer, where each node is equipped with a single-socket AMD EPYC 7763 CPU, $256$~ GB RAM, and four NVIDIA A100 (Ampere) GPUs.
Second, IPU results for the Mk2 IPUs were obtained on the ex3 supercomputer, which has a dual-socket Intel Xeon Platinum 8168 CPU connected to 16 Mk2 blades, each containing four IPU GC200s.
Third, our IPU BOW results were collected on a Paperspace Cloud instance with a dual-socket AMD EPYC 7742 CPU and $425$~ GB RAM, connected to 16 BOW IPUS in two blades. 
The experiments were compiled with native optimizations, and AVX2 was explicitly enabled with GCC 11.2.0.


\subsection{Comparison to State-of-the-Art}

To demonstrate the impact of our work, we compared our implementation to CPU -based implementations \seqan~\cite{rahn2018generic,reinert2017seqan}, ksw2~\cite{suzuki2018introducing,li2018minimap2}, libgaba~\cite{suzuki2018introducing}, and genometools~\cite{gremme2013genometools}.
All implementations were integrated into our benchmark runner program, which implements parallel processing of alignments using OpenMP~\cite{dagum1998openmp}.
Several datasets extracted from ELBA (Table~\ref{tab:datasets}) and a synthetic dataset were tested with $X \in \{5, 10, 15, 20\}$.


For our performance measurements, we define giga-cell updates per second (GCUPS) as a metric to evaluate the performance of the tools on a given data set.
Cells are defined as the number of fields in the dynamic programming matrix $S$ corresponding to the theoretical number of cells.
The time to perform a complete alignment by computing $S$ is measured by the total time $t$. 
Heuristics, such as an $X$ factor or banding, reduce the number of cells actually computed.
We define our metric as: $\textrm{GCUPS} = \frac{|\mathcal{H}|\times|\mathbf{V}|}{t}$.
The execution time was measured for our IPU implementation using the number of cycles required to compute the alignment on the device. 
The number of cycles to execute a given program is deterministic if the input and configuration parameters are identical.
Using the tile's frequency $f = 1.33 \times 10^{9}$ for the IPU Mk2 and $f = 1.85 \times 10^9$ for the IPU Bow, the total on-device execution time can be derived by $t = \nicefrac{\textrm{cycles}}{f}$.
On the GPU, the on-device execution performance was measured by measuring the kernel execution time without data transfer in LOGAN.
On the CPU, the execution time for the alignments is measured without the preparation time required for loading sequences and comparison metadata.

\subsubsection{Strong Scaling}

Using the \ECOLIX and \CELEGANS data defined in Table~\ref{tab:datasets}, we investigated the strong scaling performance of our approach, scaling a single BOW IPU to $32$ IPUs. 
Each scaling experiment was performed with graph-based partitioning (Section~\ref{sec:graph-partitioning}) enabled and disabled.
Total execution time was measured after the alignments were generated, excluding loading times for the sequences.

\subsection{Data Set}

For the standalone experiments, we generated synthetic data and extracted realistic data from the \ELBA pipeline to evaluate the performance of our own implementation as a function of certain properties of the data sets.
The datasets, including distribution characteristics, used to compare the different \xdrop implementations are listed in Table~\ref{tab:datasets}.

The distribution of our data indicates a lower sequence length of $5$~kb for \ECOLIX compared to $15$~kb for \ECOLI, and \CELEGANS.
Relative seed position on the read is relatively evenly distributed along the sequence length in our data, with a greater slope toward the center and edges in the \ECOLI and \CELEGANS datasets.
In \ECOLIX, we observed that the left and right extensions are skewed toward lower complexity alignment.
Synthetic datasets were generated with equal sequence length and fixed read similarity. Mismatches were generated by uniform-randomly mutating individual bases outside the seed position.
ELBA datasets were based on alignments generated during processing of PacBio SMRT HiFi read data from \emph{E. coli} (29x and 291x) and \emph{C. elegans} (40x) in the alignment step of the pipeline with a seed length of 17 in all datasets. ~\cite{zeni2020logan}

\subsection{Real-World Application}

For the two real pipelines \ PASTIS ( Section~\ref{sec:pastis}) and \ELBA ( Section~\ref{sec:elba}), we perform the IPU experiments on the IPU BOW system. 
Scaling to multiple IPUs was transparently enabled by setting the \texttt{ NUMBER \_IPUS} parameter of our library. 
Therefore, no further code optimization was required in either pipeline.
Both pipelines create sparse \textit{overlap} matrices to determine the sequences to be compared.
We interpret these matrices as adjacency matrices for our graph partitioning scheme presented in Section~\ref{sec:graph-partitioning}.

To compare the system performance of our systems, we focus only on the \textit{alignment} step, since the other nodes are not equipped with corresponding hardware. 
The other phases of the pipelines are not affected by the device on which the alignment step is performed.
Any speedup that one method provides over the other contributes to the speedup of the entire pipeline relative to the percentage of time that the alignment step originally took. 

\begin{figure*}[ht!]
    \centering
    \includegraphics[width=\linewidth]{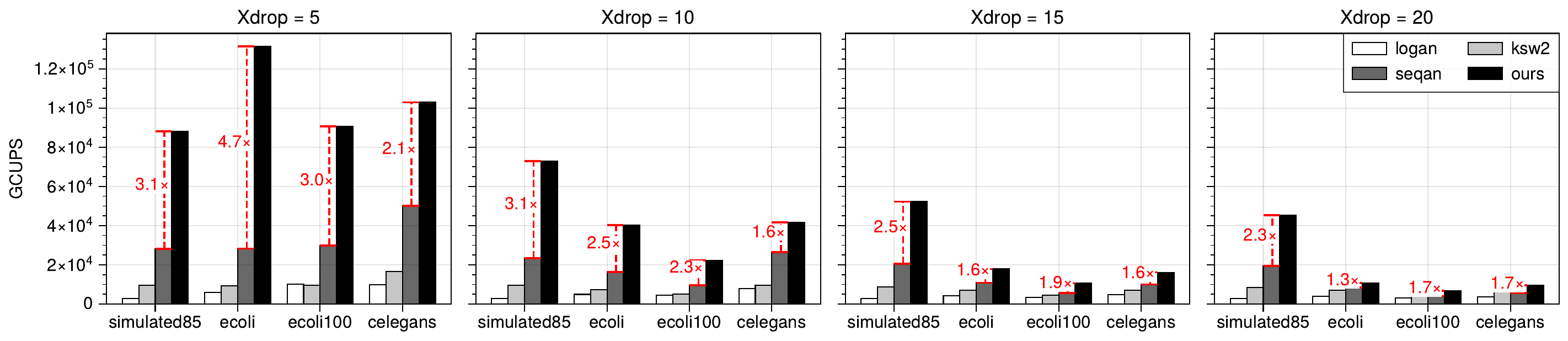}
    \vspace{-0.8cm}
    \caption{{\normalfont Normalized performance of our IPU implementation on 4 data sets (Table~\ref{tab:datasets}) in comparison to CPU implementations \seqan, ksw2 and GPU implementation LOGAN. The relative speedup to the second fasted implementation \seqan is given.}}
    \label{fig:gcups_cpu_cmp}
    \vspace{-0.3cm}
\end{figure*}

\subsubsection{PASTIS}

Our integration is based on the git-commit \texttt{fced0f2}, in which we replaced the \seqan library \xdrop alignment algorithm with our own implementation of \xdrop.
\PASTIS does not provide a \xdrop GPU algorithm because, to our knowledge, no GPU \xdrop algorithm is implemented for protein alignment.
For PASTIS, the largest dataset we could run was a uniformly subsampled protein database from the metaclust~\cite{steinegger2018clustering} dataset containing $500$~k protein sequences. 
We used an \xdrop factor of $X=49$ and a gap penalty of $-2$ and used BLOSUM62~\cite{henikoff1992amino} as our similarity matrix, as described by Selvitopi~et~al.~\cite{selvitopi2020distributed}. 
Further, we choose a $k$-mer length of 6, with two required seed matches per overlap.


\subsubsection{ELBA}

The results are based on the git-commit of the GPU branch \textit{40c1b3a}.
We used the same input data provided by Guidi et al.~\cite{guidi2022distributed} to measure performance. 
Comparisons were made with \ECOLI and \CELEGANS.
We compared the runtime of the alignment kernel in the ELBA bioinformatics pipeline.
Experiments were performed with \xdrop factors of $\{10, 15, 20\}$ and a $k$-mer length of $31$, with two required seed matches per overlap.

\section{Experimental Results}\label{sec:result}

In this section, we describe the outcomes and performance of our design choices and implementation, and compare them to the state-of-the-art.

\begin{figure}
    \centering
    \includegraphics[width=\linewidth]{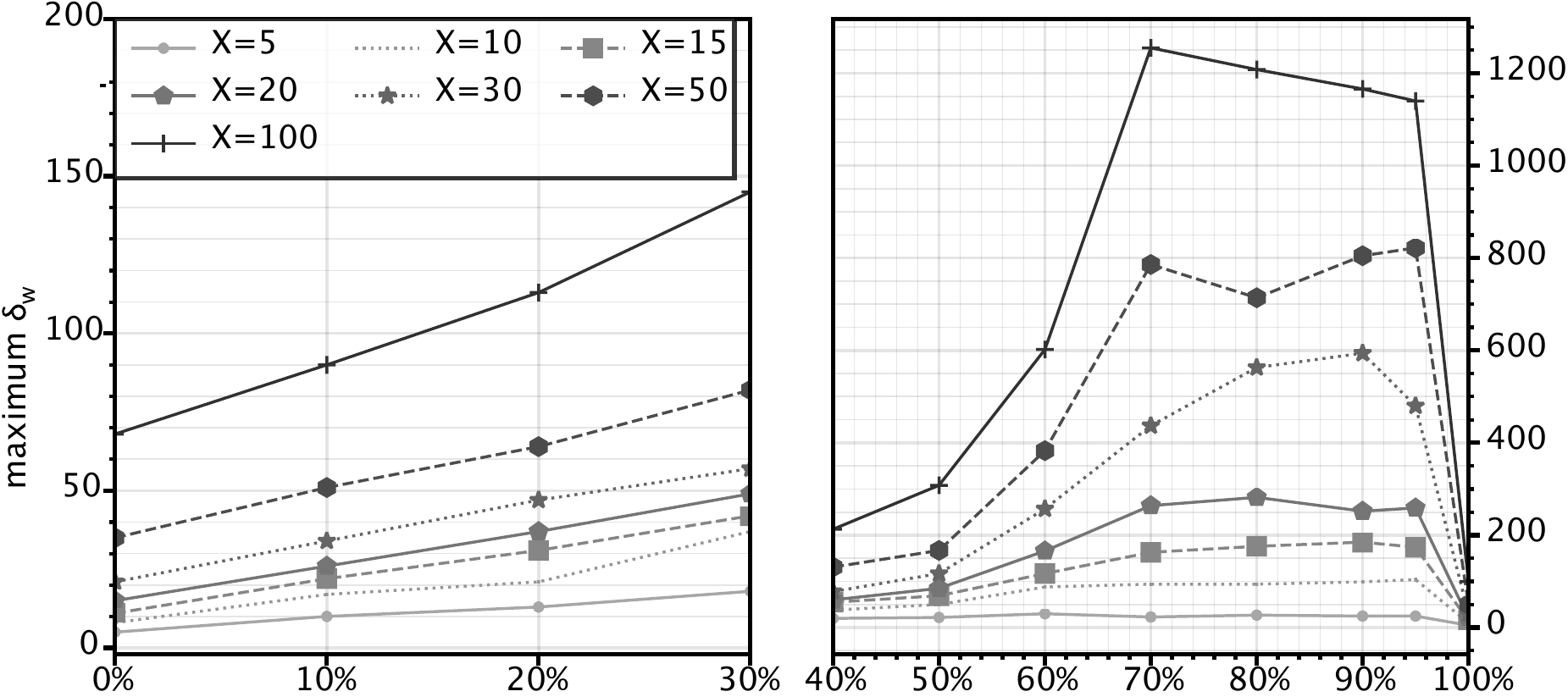}
    \vspace{-0.7cm}
    \caption{{\normalfont
    Find the maximum spread of the upper and lower pointers of the antidiagonal $\delta_w$ for error rates from $0\%$ to $100\%$ symbol mismatches with varying \xdrop values.
    }}
    \label{fig:band}
\end{figure}

\begin{figure*}
    \centering
    \includegraphics[width=\linewidth]{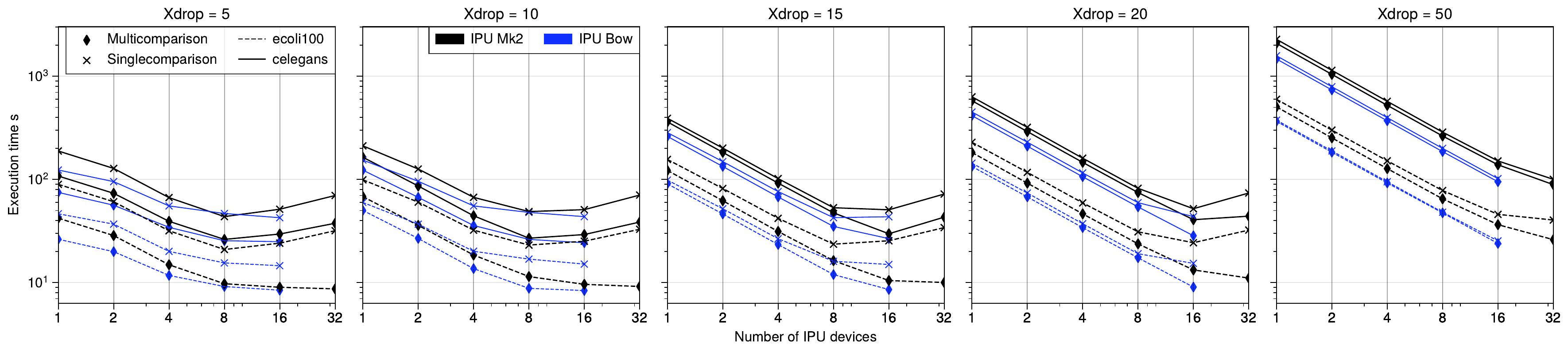}
    \vspace{-0.7cm}
    \caption{{\normalfont 
    Scaling performance measured in execution time of alignment on \ECOLIX and \CELEGANS  using 1 to 32 IPU devices. 
    \emph{Multicomparison} enables the use of a graph-based partitioning of comparisons that allows the reuse of sequences.}}
    \label{fig:ipu_scaling}
    \vspace{-0.4cm}
\end{figure*}

\subsection{Selection of $\delta_b$}
\label{sec:delta}


To test the validity of our algorithm's assumption that $\delta_w$ is significantly smaller than $\delta$, we performed experiments on different synthetic datasets with decreasing similarity rates from $100\%$ down to $0\%$ at sequence lengths of $20000$ base pairs, as shown in Figure~\ref{fig:band}.
In this empirical test, we found that the working band $\delta_w$ is smallest for perfectly matching sequences, except for sequences that are completely mismatched at a similarity of $0\%$.
For perfectly matching sequences, the computed band is small.
The highest score is always on the diagonal and the search is terminated near the diagonal based on $X$.

In the cases where the similarity decreases to $80\%$, the working band doubles for small $X$ and increases only to $13\%$ for $X=100$.
The \xdrop values studied reach a maximum bandwidth when the sequence mismatch is about $70\%$.
As the similarity decreases further, the bandwidth decreases again because the computation terminates early due to the \xdrop condition.
For fully mismatched sequences with a similarity of $0\%$, the range of computed cells is limited by $X$ to a distance from the beginning of both sequences depending on the mismatch and gap penalties.

For real-world \ECOLI data, $\delta_w$ values were $\{176, 339, 656\}$ for realistic values of $X$ of $\{10, 15, 30\}$.
Compared to the longest sequence length required for $\delta$, we can choose a $\delta_b \geq \delta_w$, which saves up to $98.2\%$ of memory for a realistic $X$ value of $15$.

\subsection{Comparison to State-of-the-Art}


Overall, the IPU implementation shows better performance, while the smallest difference was observed for \ECOLI and $X=20$ as illustrated in Figure~\ref{fig:gcups_cpu_cmp}.
The single IPU performance on the device reaches $102,844$ GCUPS with \CELEGANS at $X=5$, which is $2.05\times$ faster than that of \seqan ($50,084$ GCUPS) and $10.54\times$ faster than LOGAN ($9,761$ GCUPS) on a single GPU. 
For $X = 20$, the IPU is $1.68\times$ faster than \seqan and $2.55\times$ faster than LOGAN.


Of the CPU implementations, the implementation of \seqan \xdrop consistently performs better than ksw2 because ksw2~\cite{li2018minimap2} penalizes long gaps less, resulting in a larger search space. 
In addition, tests on an Intel Xeon Platinum 8360Y showed consistently worse performance than the AMD EPYC 7763; therefore, we have omitted the results here.
LOGAN generally does not perform well on HiFi data, and we attribute this to higher sequence similarity and a more unbalanced length distribution for smaller values of $X$.
In our implementation, we find that larger $X$ values extend the computational work to similar sequences, while dissimilar sequences continue to end early.
This leads to faster-decreasing performance for more dissimilar input sequences.
In comparison, the SIMT implementation of LOGAN gains performance over other hardware because it has a larger search space.


Figure~\ref{fig:ipu_scaling} illustrates the strong scaling from $1$ to $32$ IPU devices for $X \in \{5, 10, 15, 20, 50\}$ on \ECOLIX and \CELEGANS.
Scaling is nearly linear up to $16$ IPUs for larger $X \ge 15$ with up to $15\ times$ speedup on 16 devices on \CELEGANS ($X = 50$).
Apart from a constant speedup due to the higher clock frequency of the IPU Bow, the scaling properties do not differ significantly between the IPU Bow and IPU Mk2 systems tested.
Unless otherwise stated, the scaling results refer to the IPU Mk2 system.

Graph partitioning of comparisons allows sequences to be reused for multiple comparisons.
This reduces the amount of sequence data that must be transmitted to the IPU and increases the number of comparisons that can be performed on a single IPU tile.
For the \ECOLIX dataset, the number of batches decreases by $-52\%$ ($816$ to $387$), and for the \CELEGANS dataset, the number of batches decreases by $-44\%$ ($1,723$ to $972$).
On both data sets, this increases performance. 
Using $X = 10$, for \ECOLIX $1.46\times$ on a single device and up to $3.59\times$ using $32$ devices and for \CELEGANS $1.29\times$ on a single device and $1.83\times$ on $32$ devices.
For higher $X = 20$ and $X = 50$, scaling is linear up to $16$ and $32$ devices, both single and multiple comparisons.
or $X = 50$ using multi-comparisons is $1.18\times$ faster on the \ECOLIX dataset using $1$ devices and $1.55\times$ faster using $32$ devices.
This indicates a higher computational load per batch, allowing more IPUs to be fully utilized before the interconnect to the IPUs is saturated.

\subsection{Real World Pipelines}

The CPU and GPU results were collected on AMD EPYC 7763 nodes, while the IPU results were collected on the IPU BOW system with an AMD EPYC 7742.

\subsubsection{ELBA}

For \ECOLI, a single IPU required $7.4$~seconds to perform the alignment phase with $X=15$.
Our implementation showed good scaling up to $8$ IPUs, which reduced the alignment time to $2.2$~seconds. Scaling to more devices was restricted by the small data set.
The CPU system required $11.61$~seconds with a single node, while the GPU code ran with up to $4$ GPUs and required $52.14$~seconds in the alignment phase.
Then, we used the \CELEGANS data set, the largest data set we could run, which occupied around $400$~Gb of the host system's DRAM. 
Since the EPYC 7763 nodes have less memory, we compared the IPU with four CPU nodes and four GPU nodes.
Using a CPU setup with four nodes, we achieved $227.5, 340.7$~seconds for $X$ of $\{15, 20\}$.
GPU results were measured with a total of 16 GPUs, resulting in $1,068$~seconds for $X=15$.
The IPUs alignment phase took $255.6$ and $401.9$~seconds for $X$ of $\{15, 20\}$.
For $X \geq 15$, we observed scaling up to $16$ IPUs.
The overall alignment runtime achieved a speedup of $22.3\times$ for a cluster of $16$ GPUs and a speedup of $4.7\times$ for a four-node CPU cluster.

\subsubsection{PASTIS}

For PASTIS, we measured $44.9$~seconds for the alignment step on the CPU, while the IPU required $9.6$~seconds, which corresponds to a $4.7\times$ speedup.
Larger inputs with more than $500$~k sequences resulted in segmentation faults, making larger- scale experiments infeasible. 
This is due to the large packets combined with the $32$-bit MPI APIs, as we use a single rank with many CPU threads.
This is true for both the IPU and CPU.

\section{Related Work}\label{sec:related}
The literature on accelerating the Smith-Waterman and Needleman-Wunsch algorithms for sequence alignment is extensive~\cite{liu2013cudasw,feng2019accelerating,awan2020adept,muller2022anyseq}.
The classical sequence alignment problem without additional heuristics computes the entire dynamic programming matrix for alignment.
Other work has focused on accelerating specific tools and applications, including BLAST ~\cite{ye2017hblast,vouzis2011gpublast,ling2010design} and BWA~\cite{liu2012soap3}, both of which implement heuristic alignment strategies.
The data-intensive computational patterns in sequence alignment have led to the development of memory-centric processor architectures, including Processing in Memory (PIM) \cite{mutlu2022modern,xu2023rapidx,gupta2019rapid,zokaee2018aligner} and near-memory computing~\cite{singh2021fpgabased} systems.
Edit-distance algorithms consider only the number of changes required to convert one sequence to another, with each change, whether an insertion, deletion, or substitution, incurring the same cost. 
This much more restricted formulation of an alignment problem has similar complexity properties of $O(nm)$.
The Bitap-Algorithm~\cite{domolki1964algorithm} uses a bitmask and bitwise operations on a constrained alphabet to compute the edit distance between a pattern and a queried string. 
It was adapted with a greedy windowing heuristic for long sequences and parallelized execution in GenASM and Scrooge~\cite{cali2020genasm,lindegger2023scrooge}, reducing the alignment complexity to $O(n+m)$, while also allowing for non-optimal alignment results.

The \xdrop algorithm for alignment has rarely been the target of hardware acceleration work.
Recent efforts include GPU~\cite{zeni2020logan} and FPGA (Field Programmable Gate Array)~\cite{zeni2021importance}, which surpassed the state-of-the-art CPU implementations at the time.


\section{Conclusion}\label{sec:conclusion}
The processing power of modern CPUs has far outpaced the speed improvement of persistent and random access memory (DRAM), widening the gap between memory and processor performance.
This discrepancy is masked by a highly hierarchical cache system in traditional CPUs.
The Graphcore IPU's single level of large, low-latency SRAM reflects the recent trend toward shifting computation to memory, in both PIM and near-memory computing approaches.

In this work, we implement the \xdrop sequence alignment algorithm on the Graphcore IPU.
A massively parallel MIMD AI accelerator with a single level of low latency SRAM for storage.
Our contributions include algorithmic updates to the \xdrop algorithm to adapt it to the memory-constrained IPU architecture.
Our dynamic band restriction algorithm reduces memory usage without compromising alignment computation with real data.
Our formulation of graph-based sequence partitioning enables the reuse of sequences in many-to-many sequence alignment settings common to genome assembly and protein cluster pipelines.

Our implementation of \xdrop sequence alignment outperforms current state-of-the-art implementations on CPU and GPU for both DNA and protein alignment for realistic $X$ values.
In addition, we demonstrate near-linear strong scaling properties on common IPU host configurations.
In two real-world pipelines, ELBA and PASTIS, we show significant speedup using our IPU implementation as the algorithm for the \xdrop aligner.

Finally, we note that the low bandwidth of host-device communication and the rigidity of the BSP paradigm, as well as the lack of atomic operators for thread-level cooperative multitasking, are the major limitations of the Graphcore IPU system.
Our implementation mitigates these issues, but future SRAM-based architecture should be improved to enable the widespread use of SRAM-based computing for more data-intensive computation.

In summary, the IPU has significant potential for accelerating irregular computations where low-level parallelism is difficult to exploit on highly instruction-parallel architectures such as GPUs.



\bibliography{bibliography}
\balance

\end{document}